\documentclass[trackchanges,twocolumn]{aastex701}
\usepackage{amsmath}

\begin{document}

\title{Three-dimensional reddening maps of the Magellanic Clouds constructed by RR Lyrae stars}

\author[orcid=0009-0007-4233-5334,gname=Shun-Xuan,sname=He]{Shun-Xuan He}
\affiliation{School of Physics and Astronomy, Beijing Normal University, Beijing 100875, People's Republic of China}
\affiliation{Institute for Frontiers in Astronomy and Astrophysics, Beijing Normal University, Beijing 102206, People's Republic of China}
\email{202531101090@mail.bnu.edu.cn}

\author[orcid=0000-0003-3250-2876,gname=Yang,sname=Huang]{Yang Huang}
\affiliation{University of Chinese Academy of Sciences, Beijing 100049, People's Republic of China}
\affiliation{National Astronomical Observatories, Chinese Academy of Sciences, Beijing 100101, People's Republic of China}
\email[show]{huangyang@ucas.ac.cn}
\correspondingauthor{Yang Huang}

\begin{abstract}

We present the first three-dimensional reddening maps of the Large and Small Magellanic Clouds (LMC and SMC) constructed using fundamental-mode RR Lyrae stars from the OGLE-IV survey. By applying a period-amplitude–color relation and a period–luminosity–metallicity calibration in the OGLE photometric system, we derive intrinsic colors, color excess $E(V-I)$, and photometric distances for more than 20,000 RRab stars in the LMC and 3,000 in the SMC. Spatial variations in reddening are modeled using an adaptive quadtree scheme, where robust reddening–distance relations are fit within each partition and distances are iteratively updated to achieve self-consistency. The resulting maps reveal resolved dust structures across both galaxies, including steep reddening gradients in the central LMC and flatter profiles in the SMC. The construction of the three-dimensional reddening maps further reveals that high-extinction regions exhibit reddening behavior inconsistent with a uniform extinction law, implying localized variations in dust properties. The final maps comprise 205 partitions for the LMC and 67 partitions for the SMC, and are released together with a Python-based query tool and GeoJSON data products. These 3D maps provide a foundation for distance-dependent reddening corrections and for probing the structure and physical conditions of the Magellanic interstellar medium, and future high-precision and cadence RR Lyrae sample from Gaia DR4 will support higher-resolution mapping and deeper exploration of dust substructure.

\end{abstract}



\section{Introduction} 

Interstellar dust is a fundamental component of galaxies, shaping their physical, chemical, and observational properties. It regulates the thermal and chemical state of the interstellar medium, provides sites for molecule formation, and influences star formation and radiative transfer \citep{Draine2003,Galliano2022}. Dust extinction further alters the observed colors and luminosities of astronomical sources \citep{Trumpler1930}, making accurate extinction measurements essential for deriving intrinsic stellar parameters, determining reliable distances, and reconstructing galactic star formation and evolution histories \citep{Galliano2018,Salim2020}.

In the Milky Way, both two-dimensional (2D) and three-dimensional (3D) extinction maps have been developed. The seminal 2D all-sky maps, derived from far-infrared dust emission \citep{SFD1998}, provide the integrated line-of-sight extinction and have served as a fundamental tool for Galactic foreground reddening correction. Their methodology and accuracy were later refined using data from the Planck \citep{Planck2014,Meisner2015,Planck2016}. More recent 3D maps combine large-scale photometric surveys with distance estimates to resolve dust structures along the line of sight (e.g., \citealt{Lallement3d2018,Green3d2019,Chen3d2019,Guo3d2021,Wang2025}). These studies have greatly improved our understanding of Galactic dust geometry and enabled precise dereddening of stellar populations. Comparable 3D mapping for external galaxies, however, remains limited.

The Magellanic Clouds (MCs), the Large and Small Magellanic Clouds (LMC and SMC) provide a valuable opportunity to extend such studies beyond the Milky Way. Their proximity, low metallicities, and diverse interstellar environments make them ideal for investigating dust formation and evolution under conditions distinct from those in our Galaxy \citep{Rubio2023}. Accurate extinction maps of the MCs are also crucial for calibrating the extragalactic distance scale \citep{Grijs2014,Riess2019,Freedman2020} and reconstructing spatially resolved star formation histories \citep{Meixner2014,Mazzi2021}. Over the past two decades, extensive 2D reddening maps have been derived from Red Clump stars (e.g., \citealt{Subramaniam2005,Subramanian2012,Haschke2011,Tatton2013,Choi2018,Gorski2020,Skowron2021,Nataf2021}), Cepheids (e.g., \citealt{Inno2016,Joshi2019}), RR Lyrae stars (e.g., \citealt{Pejcha2009,Haschke2011,Deb2017,Muraveva2018,Cusano2021}), and spectral energy distribution fitting of stars or background galaxies (e.g., \citealt{Bell2020,Bell2022,Chen2022}). While these studies reveal that dust in the MCs is highly structured, 2D maps integrate extinction along the line of sight, potentially introducing systematic biases where stars and dust are mixed. A three-dimensional approach, resolving extinction as a function of distance, is therefore necessary to uncover the internal geometry of the MCs and provide reliable, distance-dependent reddening corrections.

RR Lyrae (RRL) stars are ideal tracers for constructing 3D extinction maps, as they are old, metal-poor horizontal-branch pulsators with well-defined period–luminosity–metallicity (PMZ) relations in optical and infrared bands, serving as robust distance indicators (e.g., \citealt{Muraveva2018rr,Bhardwaj2022,Li2023,He2025}). The fourth phase of the Optical Gravitational Lensing Experiment (OGLE-IV; \citealt{Udalski2015}) offers a homogeneous catalog of RR Lyrae stars across both Clouds, with deep V- and I-band light curves and high cadence \citep{Soszynski2016}. These data enable precise determination of stellar distances and color excesses, providing the foundation for detailed 3D extinction mapping.

In this article, we present the first 3D reddening maps of the LMC and SMC based on OGLE-IV RR Lyrae stars. By combining distance and color-excess estimates, we map the spatial variation of reddening throughout the Clouds. Section 2 describes the sample selection, distance and color-excess estimation, and map construction methods and talk about some caveats of our model. Section 3 presents the resulting 3D reddening maps, highlights their main features, and discusses the assumptions and limitations inherent to the modeling. Section 4 summarizes the results and discusses future applications.

\section{Methods}\label{sec:samp}

This section presents the estimation of RR Lyrae physical parameters, including color excess and distance, which are crucial for constructing reddening maps, followed by the procedure for building the maps of the MCs.

\subsection{Color Excess and Distance Estimation for the OGLE RR Lyrae Sample}

\subsubsection{Color Excess $E(V-I)$}

The color excess of a star is typically determined as the difference between its observed and intrinsic colors. The intrinsic color can be estimated in two ways: (1) by subtracting the absolute magnitudes predicted from the period–luminosity–metallicity (PMZ) relations in two different bands, or (2) by applying a direct period–color–metallicity (PCZ) relation. Both methods, however, are often limited by the accuracy of the metallicity estimates. An alternative approach is provided by the empirical period–amplitude–color (PAC) relation for fundamental-mode RR Lyrae stars (RRab) introduced by \citet{Piersimoni2002}, which effectively removes the metallicity dependence in determining $(V - I)_0$, as shown in Equation (1). This relation achieves relatively high precision and has been adopted for extinction estimation in the dedicated RR Lyrae processing pipelines of Gaia DR2 and DR3 missions \citep{Clementini2019,Clementini2023}.

\begin{equation}
\begin{aligned}[b]
(V-I)_{0} = & (0.65\pm0.02) - (0.07\pm0.01)\times\mathrm{Amp}(V) \\
            & + (0.36\pm0.06)\times \log(P) \\
            (\sigma = 0.02)
\end{aligned}
\end{equation}

The OGLE-IV project provides photometric data for more than 32,000 RRab stars with both $V$- and $I$-band light curves in the MCs, identified by IDs containing 'LMC' or 'SMC' (hereafter referred to as the \textbf{MC\_BASE} sample) . The observed color $(V-I)$ of them is derived from mean magnitudes obtained through light-curve decomposition. Combining these with the intrinsic color $(V-I)_0$ from the above PAC relation allows for a straightforward estimation of the color excess $E(V-I)$, which in turn facilitates the construction of the reddening map for the MCs. 

However, since OGLE provides a significantly higher number of observations in the $I$ band than in the $V$ band, it is more practical to convert the $V$-band amplitude in Equation (1) into the $I$ band to improve the estimation accuracy. We therefore selected a calibration subsample of OGLE RRab stars with sufficient observational epochs ($\geq 30$) and full phase coverage ($\geq 0.9$; one minus the maximum of adjacent phase difference) in both $V$ and $I$ bands. Their amplitudes and corresponding uncertainties were derived using a Fourier decomposition method \citep{Dekany2021,He2025}. A Bayesian linear fit accounting for amplitude uncertainties yields the following calibration relation:

\begin{equation}
\begin{aligned}[b]
\mathrm{Amp}(V) = & (1.598 \pm 0.001) \times \mathrm{Amp}(I) + (0.0005 \pm 0.0004) \\
(\sigma = 0.035)\text{.}
\end{aligned}
\end{equation}
Combining Equations (1) and (2), we obtain a revised PAC relation expressed in terms of the $I$-band amplitude, suitable for the \textbf{MC\_BASE} sample:

\begin{equation}
\begin{aligned}[b]
(V-I)_{0} = & (0.65\pm0.02) - (0.112\pm0.016)\times\mathrm{Amp}(I) \\
            & + (0.36\pm0.06)\times \log(P) \\
            (\sigma = 0.02)\text{.}
\end{aligned}
\end{equation}
The reddening is then calculated as: $E(V-I)=(V-I)-(V-I)_0$. The uncertainty in $E(V-I)$ is dominated by the error in the intrinsic color estimate, derived from error propagation of the PAC relation parameters combined with the intrinsic scatter of the relation. Our $E(V-I)$ measurements show good consistency with the values from the 2D red clump–based reddening map of \citet{Skowron2021}, exhibiting a mean offset of only 0.03 mag and a scatter of 0.05 mag. This close agreement supports the reliability of our reddening estimates.

\subsubsection{Distance Estimation}

We first estimated the absolute magnitude $M_I$ of the \textbf{MC\_BASE} sample using the PMZ relation in the $I$ band provided by \citet{Prudil2024}, which is calibrated in the OGLE photometric system. The period term was directly adopted from the OGLE catalog. For the metallicity term, we derived photometric metallicities based on the $P$–$\phi_{31}$–$A_2$–[Fe/H] relation calibrated by \citet{Dekany2021}. Here, $A_2$ and $\phi_{31}$ (defined as $\phi_3 - 3\phi_1$) represent the second Fourier amplitude and the phase difference, respectively, obtained through Fourier decomposition in $I$ band. The resulting photometric metallicities are on the same metallicity scale as that used in the adopted PMZ relation. The uncertainties in photometric metallicities were rigorously estimated through 1000 Monte Carlo simulations, incorporating both the intrinsic scatter of the calibration relation and the measurement errors in the Fourier parameters.  

A second key ingredient in the distance determination is the extinction $A_I$, which depends on the extinction coefficient $R_I$ and the color excess $E(V-I)$. Using our own derived $E(V-I)$ values at this stage would couple reddening and distance, removing their independence and potentially producing artificial correlations in the reddening–distance relation. To avoid this, we used the external two-dimensional map $E(V-I)_{\mathrm{ini}}$ from \citet{Skowron2021} to compute the initial, independently determined distances $d_\mathrm{ini}$. Together with our PAC-based color excesses, these distances provide the initial configuration in reddening–distance space. In regions where the initial distribution of $E(V-I)$ and distance suggested a genuine reddening–distance trend, the distances were later refined through an iterative procedure that reconciles the two independent reddening estimates, as described in Section~\ref{sec:const_3dmaps}.

For the extinction coefficients, we adopted $R_V = 3.40$ and $2.53$ for the LMC and SMC, respectively, as estimated from red supergiants and classical Cepheids \citep{Wang2023}. Using the corresponding extinction laws at the effective wavelength of 798\,nm, we derived $A_I/A_V$ ratios and obtained $R_I = 1.35$ for the LMC and $R_I = 1.10$ for the SMC. The extinction was then calculated as $A_I = R_I \times E(V-I)_{\mathrm{ini}}$ and the distance $d_\mathrm{ini}$ of the \textbf{MC\_BASE} sample was subsequently derived. Distance uncertainties were comprehensively estimated by propagating errors from absolute magnitudes, apparent magnitudes and extinction values, ensuring robust uncertainty characterization for subsequent 3D reddening mapping.

\subsection{Construction of the 3D Reddening Maps}\label{sec:const_3dmaps}

With both color excess and distance estimates derived for the RR Lyrae sample, we proceeded to construct the three-dimensional reddening maps of the MCs. To ensure the reliability of the reddening–distance relation, we first refined the dataset by applying a series of quality cuts to the \textbf{MC\_BASE} sample, yielding the final working sample, \textbf{MC\_FINAL}.

The selection criteria are summarized as follows:
\begin{itemize}
\item A $3\sigma$ clipping was applied to the distance estimates to remove outliers, which are likely non-members including foreground and background stars.
\item Relative distance uncertainties were required to be smaller than 12\% to ensure reliable distance determinations.
\item The uncertainty in color excess, $\sigma_{E(V-I)}$, was limited to less than 0.05\,mag to guarantee accurate extinction estimates.
\item Only stars with physically reasonable color excess values in the range $0 \leq E(V-I) \leq 0.8$\,mag were retained, excluding unphysical negative values and abnormally high extinctions that may have lower accurcy.
\end{itemize}

After applying these selection criteria, the final high-quality sample comprised approximately 20,309 RRab stars in the LMC and 3,308 in the SMC.

\subsubsection{Adaptive Spatial Partitioning Framework}

\begin{figure*}[!htbp]
  \centering
  \includegraphics[width=0.98\textwidth,keepaspectratio]{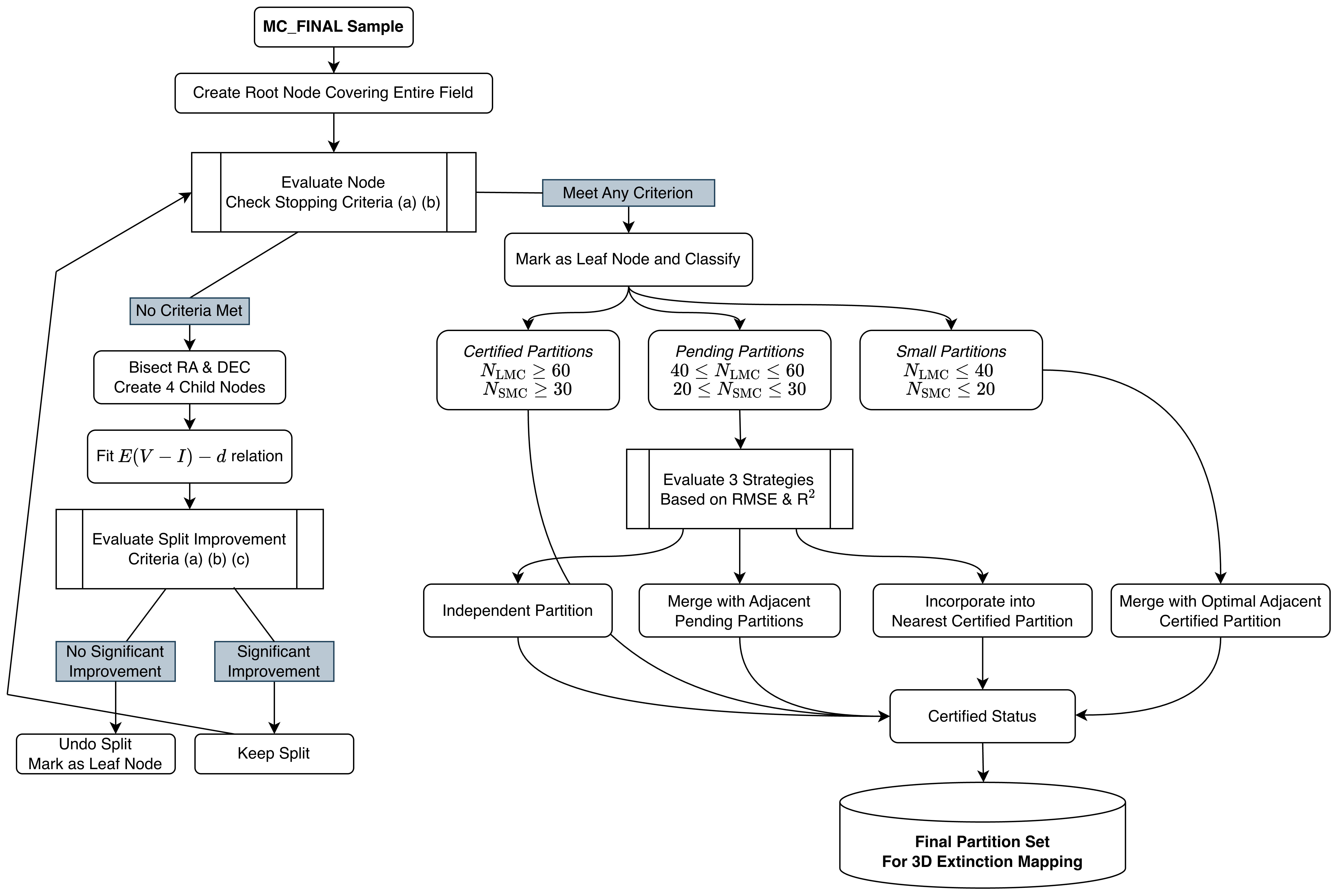}
  \caption{Flow diagram illustrating the adaptive partitioning method.}
  \label{fig:partition framework}
\end{figure*}

To capture the spatial variation of reddening while maintaining sufficient stellar statistics, we implemented an adaptive quadtree partitioning algorithm that divides the \textbf{MC\_FINAL} sample into spatial cells in equatorial coordinates. The overall workflow is illustrated in Figure~\ref{fig:partition framework}. The algorithm begins by assigning all stars to a root node encompassing the entire field of view, and proceeds recursively through breadth-first traversal. At each node, subdivision is evaluated according to two stopping criteria: (1) an insufficient number of stars for reliable fitting ($N_{\mathrm{min}} = 60$ for the LMC and 30 for the SMC), and (2) an angular size below a minimum threshold (0.01\,deg), which avoids over-subdivision into cells too small to yield robust or spatially coherent reddening trends. When neither condition is met, the node is divided into four subcells by bisecting both the R.A. and Dec. directions.

For each child node, the $E(V-I)$–$d_\mathrm{ini}$ relation is determined using a robust linear regression with iterative $3\sigma$ clipping. This linear form adequately describes the observed reddening–distance correction within the limited spatial extent of each partition, where the dust distribution can be reasonably approximated as uniform and higher-order variations are not warranted by the data precision. Subdivision is accepted only when it provides meaningful improvement, following three criteria: (a) the current node does not trigger early stopping, (b) at least one child node remains eligible for further subdivision, and (c) either the current node exhibits poor fit quality , with a root-mean-square Error (RMSE) exceeding 0.03\,mag for LMC (0.02\,mag for SMC) or at least one child achieves a significant improvement ($\geq$10\% reduction in RMSE). Nodes satisfying these conditions are placed back into the processing queue for continued subdivision, while the rest are retained as terminal leaves, resulting in a hierarchical structure of variable spatial resolution.

The resulting leaf nodes are then categorized according to their stellar counts relative to $N_{\mathrm{min}}$. Nodes exceeding this limit and exhibiting satisfactory fit quality are designated as \textit{certified partitions}. Nodes below the threshold, which originate from parents that satisfied the division criteria, are processed using an adaptive merging strategy to address low tracer density while maintaining spatial coherence. These are further divided by a secondary threshold, $N_{\mathrm{min,fit}}$ (40 for the LMC and 20 for the SMC), into two subgroups: \textit{pending partitions} ($N_{\mathrm{min,fit}} \leq N_{\mathrm{star}} < N_{\mathrm{min}}$) and \textit{small partitions} ($N_{\mathrm{star}} < N_{\mathrm{min,fit}}$).

For pending partitions, three alternative treatments are evaluated: (1) retaining them as independent partitions, (2) merging with adjacent pending partitions, or (3) incorporating them into neighboring certified partitions. The optimal configuration is determined based on improvements in both RMSE and the coefficient of determination ($R^2$) relative to independent fitting. Pending partitions are promoted to certified status once the optimal configuration is identified. Small partitions are automatically merged with adjacent certified partitions that yield the smallest post-merging RMSE after refitting. The final set of partitions is then established.

This hierarchical approach ensures that all available RR~Lyrae tracers contribute meaningfully to the construction of the reddening map, while preserving statistical reliability through adaptive spatial resolution and rigorous evaluation of the reddening–distance relations within each partition.

\subsubsection{Iterative Fitting of the Reddening–Distance Relation}\label{sec:iter_fit}

After defining the spatial partitions, we implemented an iterative fitting procedure to refine the reddening–distance relation within each cell. This process was designed to reconcile the inconsistencies between two independent estimates of color excess: the initial photometric distances $d_\mathrm{ini}$, derived from the integrated line-of-sight reddening $E(V-I)_\mathrm{ini}$ from an external map, and the color excess $E(V-I)$ derived from the PAC relation. By progressively replacing the 2D reddening prior with data-driven corrections, this iterative process yields a self-consistent 3D reddening model in which distances and reddening estimates converge simultaneously across all spatial partitions.

The iteration begins with the initial photometric distances $d_\mathrm{ini}$ derived from the PMZ relation. Within each partition, we perform a robust linear regression between the color excess and distance, expressed as

\begin{equation}
E(V-I) = \alpha \cdot d_\mathrm{ini} + \beta\text{,}
\end{equation}
where $\alpha$ quantifies the reddening gradient along sightlines passing through the MCs and $\beta$ denotes the intercept. The fitting procedure adopts an iterative $3\sigma$-clipping algorithm to remove outliers, assigning weights according to the uncertainties in $E(V-I)$. Only stars retained after clipping contribute to the final regression, ensuring robustness against anomalous data points.

Once the linear relation is established, stellar distances are updated according to

\begin{equation}
d_{\mathrm{new}} = d_\mathrm{ini} \times 10^{R_I \cdot (E(V-I)_{\mathrm{ini}} - E(V-I)_{\mathrm{model}})/5}\text{,}
\end{equation}
where $E(V-I)_\mathrm{ini}$ is the initial color excess from \citet{Skowron2021} used to compute $d_\mathrm{ini}$, $E(V-I)_\mathrm{model}$ is the model-predicted value from the fitted relation at current distance, and $R_I$ is fixed to the same value adopted in Section~2.1.2. The iteration alternates between refitting the $E(V-I)$–$d_\mathrm{new}$ relation using updated distances and recalculating distances based on the new fits. Convergence is assessed independently for each partition by requiring the relative change in distance between successive iterations to be below 0.001 for all stars within the partition. A maximum of 50 iterations is imposed; partitions that do not meet this criterion are terminated and examined separately.

Most partitions converged smoothly within 10–20 iterations. However, a small fraction of regions in the LMC exhibited oscillatory behavior, where distance estimates fluctuated by several to nearly twenty percent despite stable regression coefficients. These partitions are primarily located in regions of high extinction, characterized by elevated $E(V-I)$ and $A_V$ values. Comparison with the two-dimensional $R_V$ map from \citet{Zhang2025} suggests that such regions typically exhibit lower $R_V$ (and correspondingly lower $R_I$) than the adopted value. We therefore attribute the oscillation to local discrepancies between the fixed $R_I$ and the true dust properties. In these cases, the assumption of $R_I = 1.35$ resulted in overcorrections during distance updates, leading the solutions to oscillate around their equilibrium.

To mitigate this issue, we introduced an adaptive adjustment of $R_I$ for non-converging partitions. For stars that remained divergent after the maximum iteration count, $R_I$ was reduced incrementally by 5\% per adjustment round, with a lower limit of $R_I = 0.8$ to maintain physical plausibility. After each adjustment, the extinction $A_I$ and corresponding distances were recomputed, and the entire fitting procedure was rerun. This process was repeated until convergence was achieved.

\begin{figure*}[!htbp]
  \centering
  
  \includegraphics[width=0.95\textwidth,keepaspectratio]{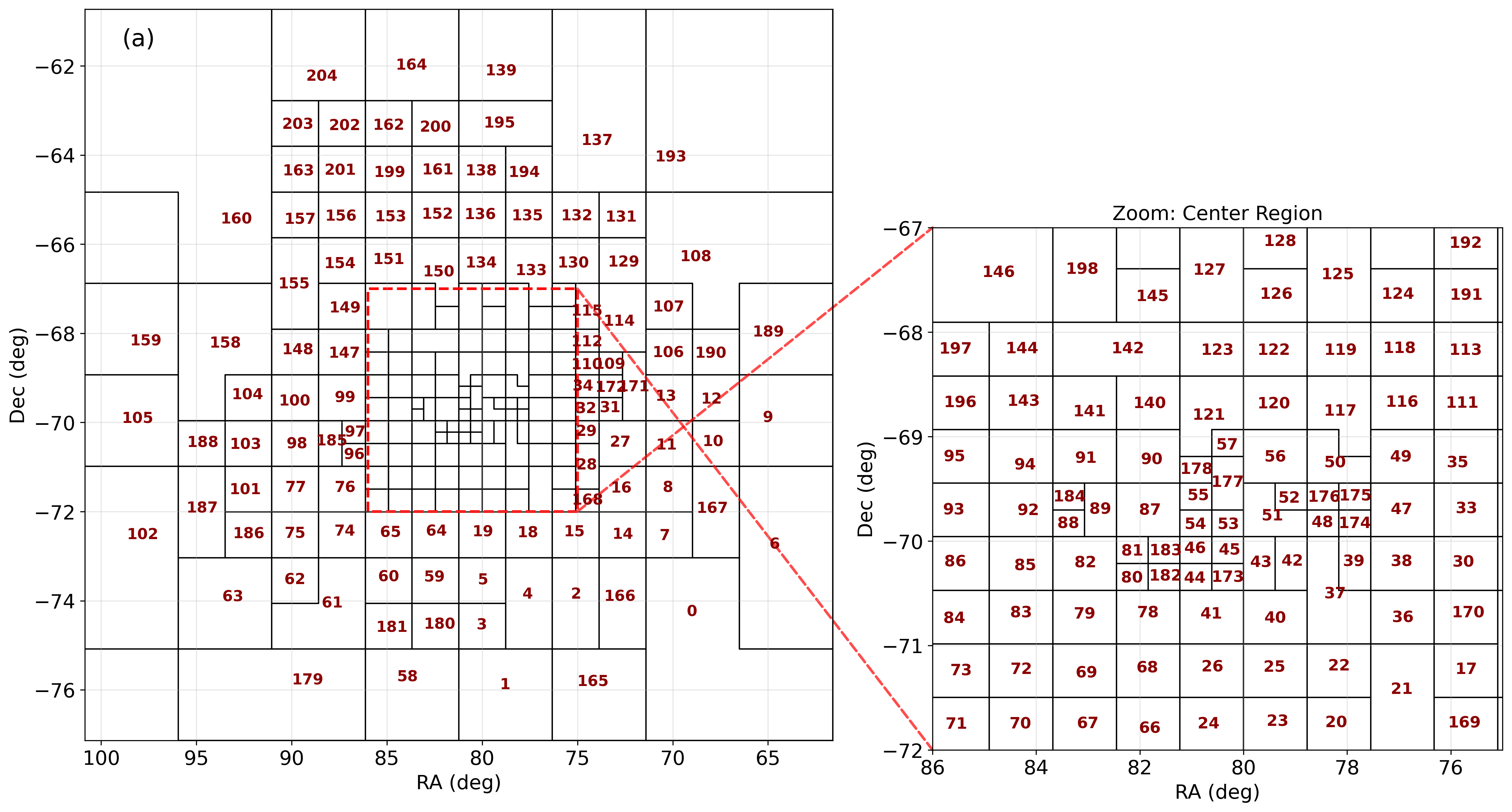}
  \vspace{0.2cm}
  
  \includegraphics[width=0.49\textwidth,keepaspectratio]{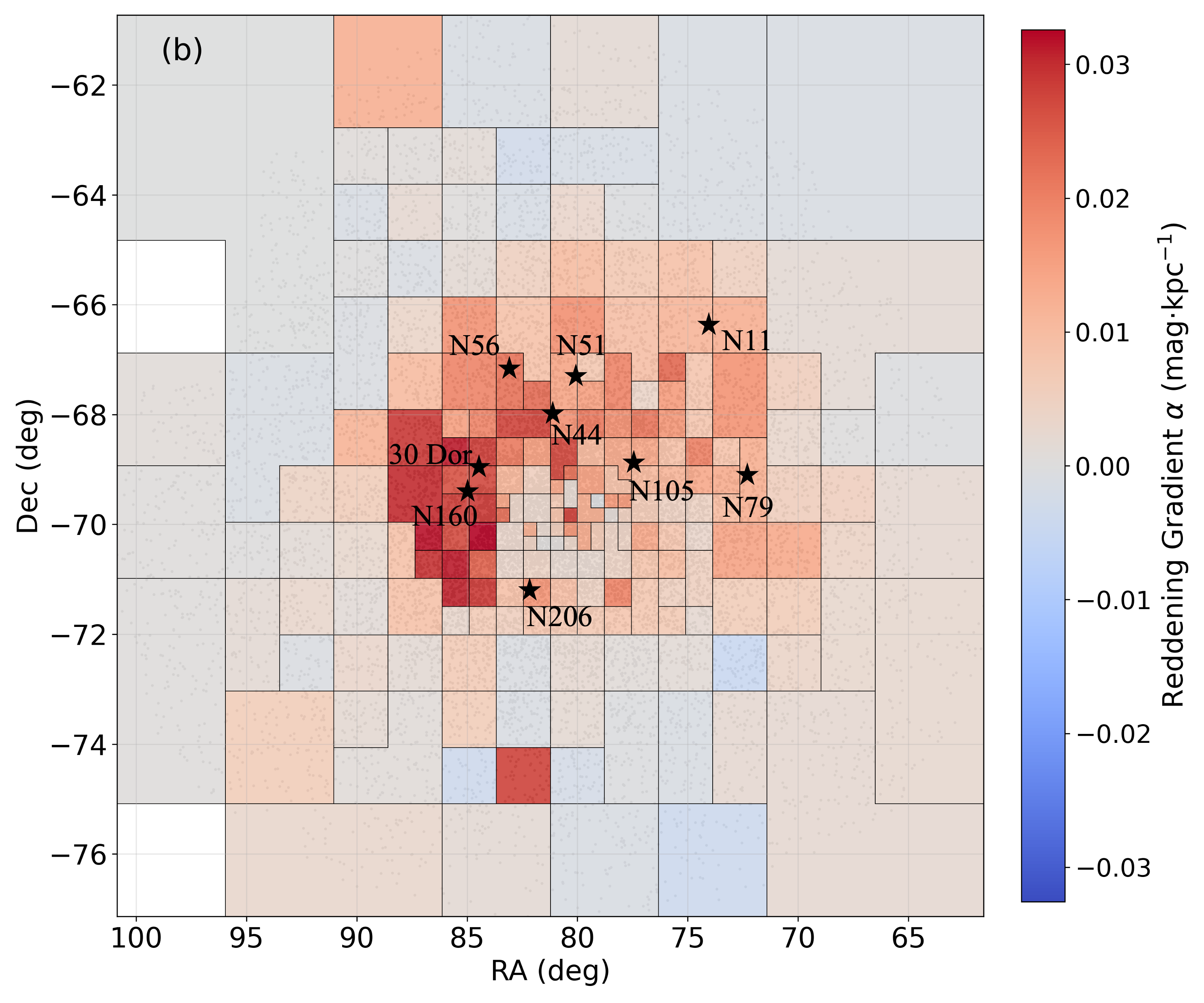}
  \hfill
  \includegraphics[width=0.49\textwidth,keepaspectratio]{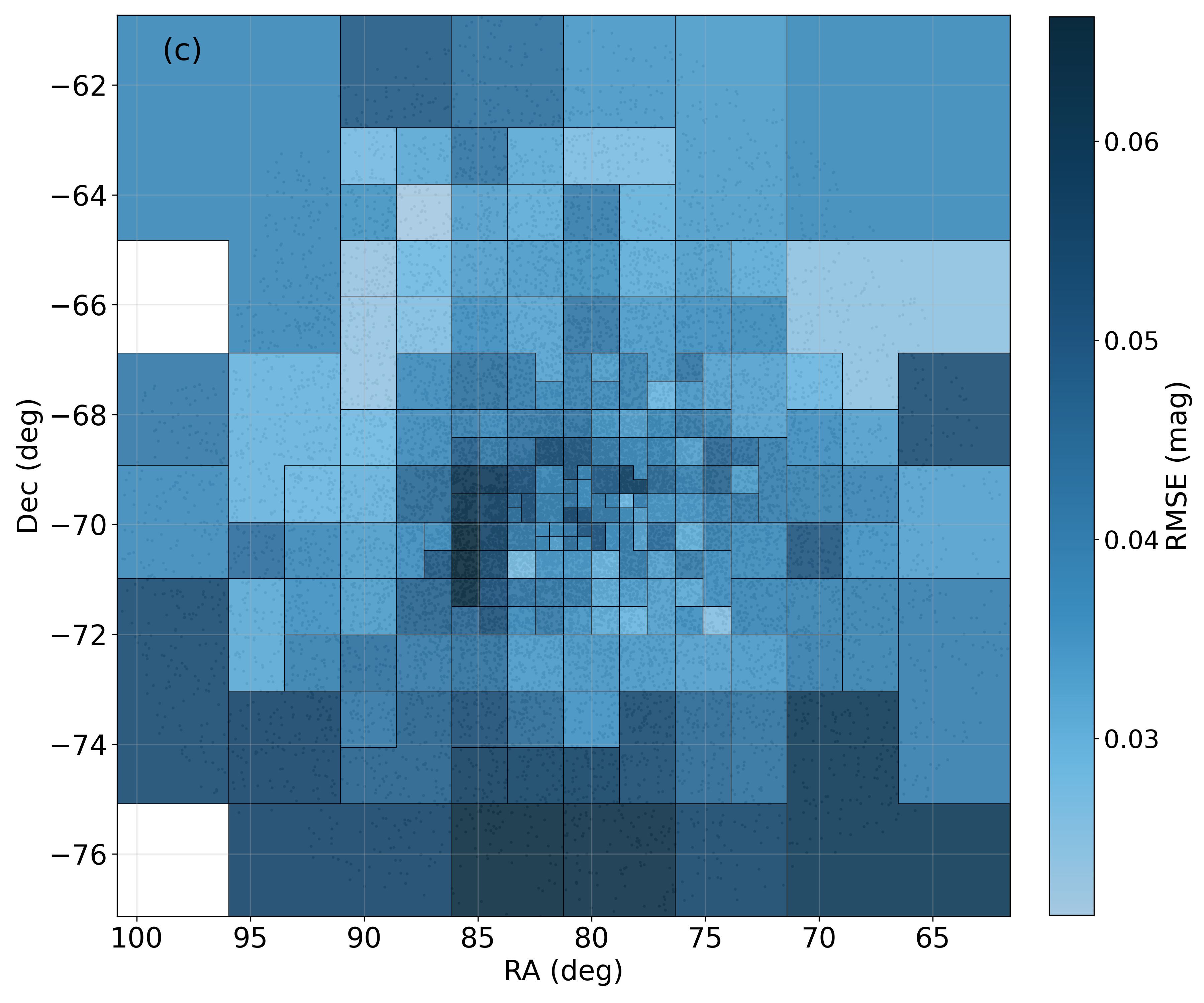}
  
  \vspace{0.2cm}
  
  \caption{Overview of the spatial subdivision and fitting performance of the LMC reddening map. Panel (a): Spatial partitions of the LMC, with each subregion labeled by its red ID; a zoomed-in view of the central area is provided for clarity. Panel (b): Reddening gradients from the linear reddening–distance fits, color-coded by gradient magnitude. Star symbols mark representative star-forming regions, labeled according to the catalog of \citet{Henize1956}. Panel (c): RMSE of the fits in each partition, where darker colors indicate larger residuals.}
  \label{fig:LMC_fit_result}
\end{figure*}

This adaptive scheme effectively suppressed oscillations by aligning the assumed extinction ratio with local dust conditions, while preserving the established spatial partitioning. The final converged solutions provide self-consistent reddening–distance relations across all cells, forming the foundation of our three-dimensional reddening maps.


\begin{figure*}[!htbp]
  \centering
  
  \includegraphics[width=0.5\textwidth,keepaspectratio]{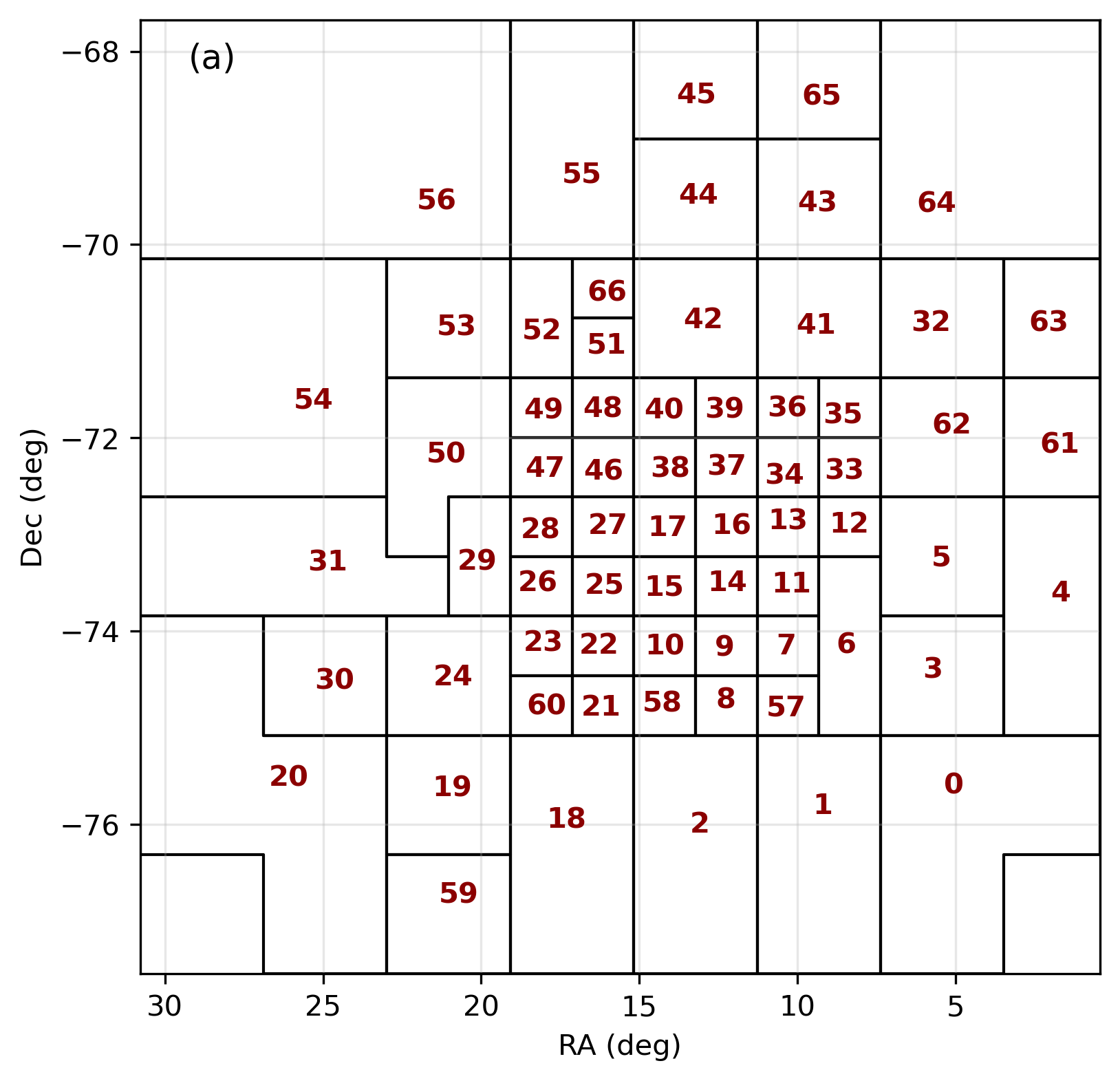}
  \vspace{0.2cm}
  
  \includegraphics[width=0.49\textwidth,keepaspectratio]{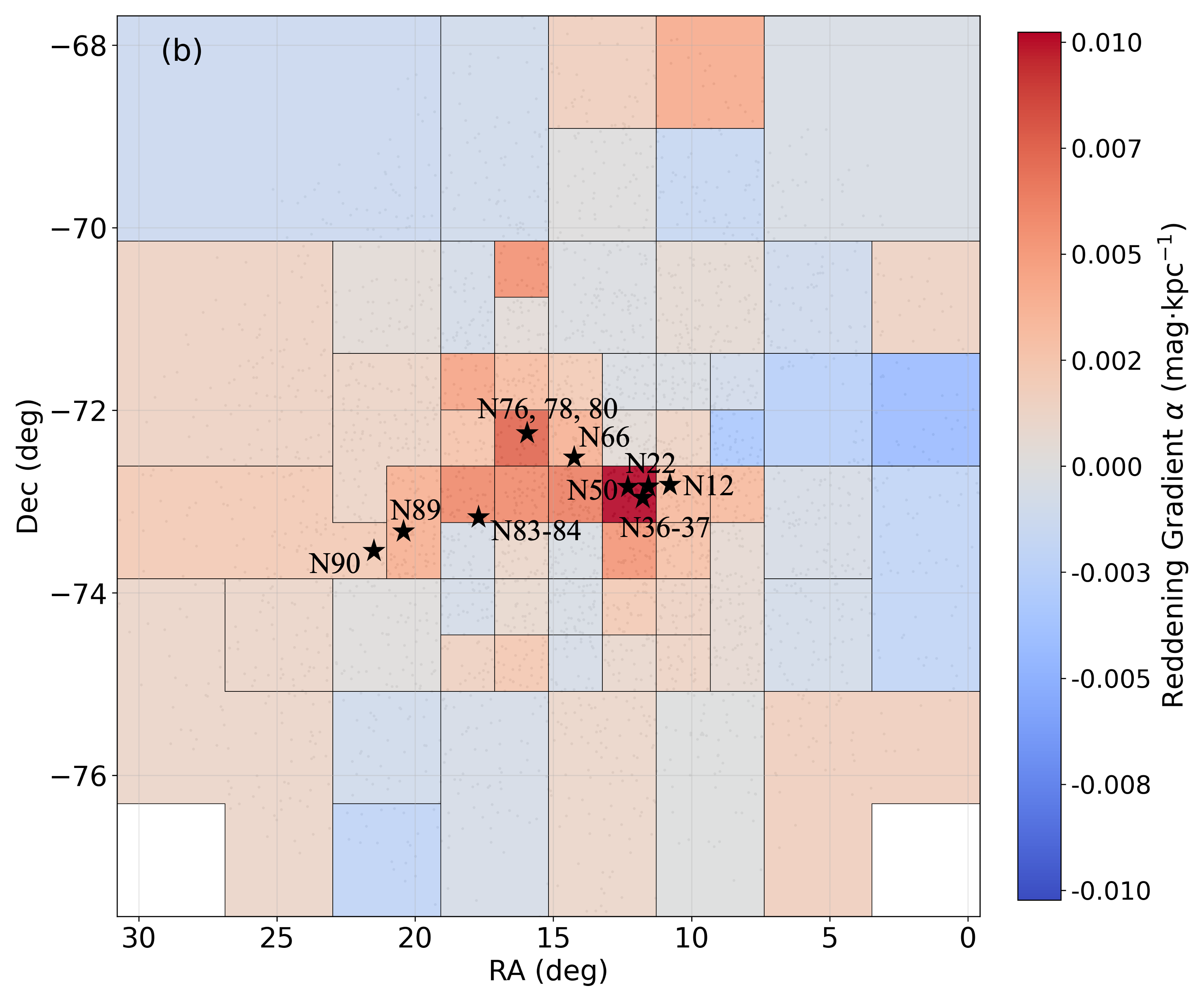}
  \hfill
  \includegraphics[width=0.49\textwidth,keepaspectratio]{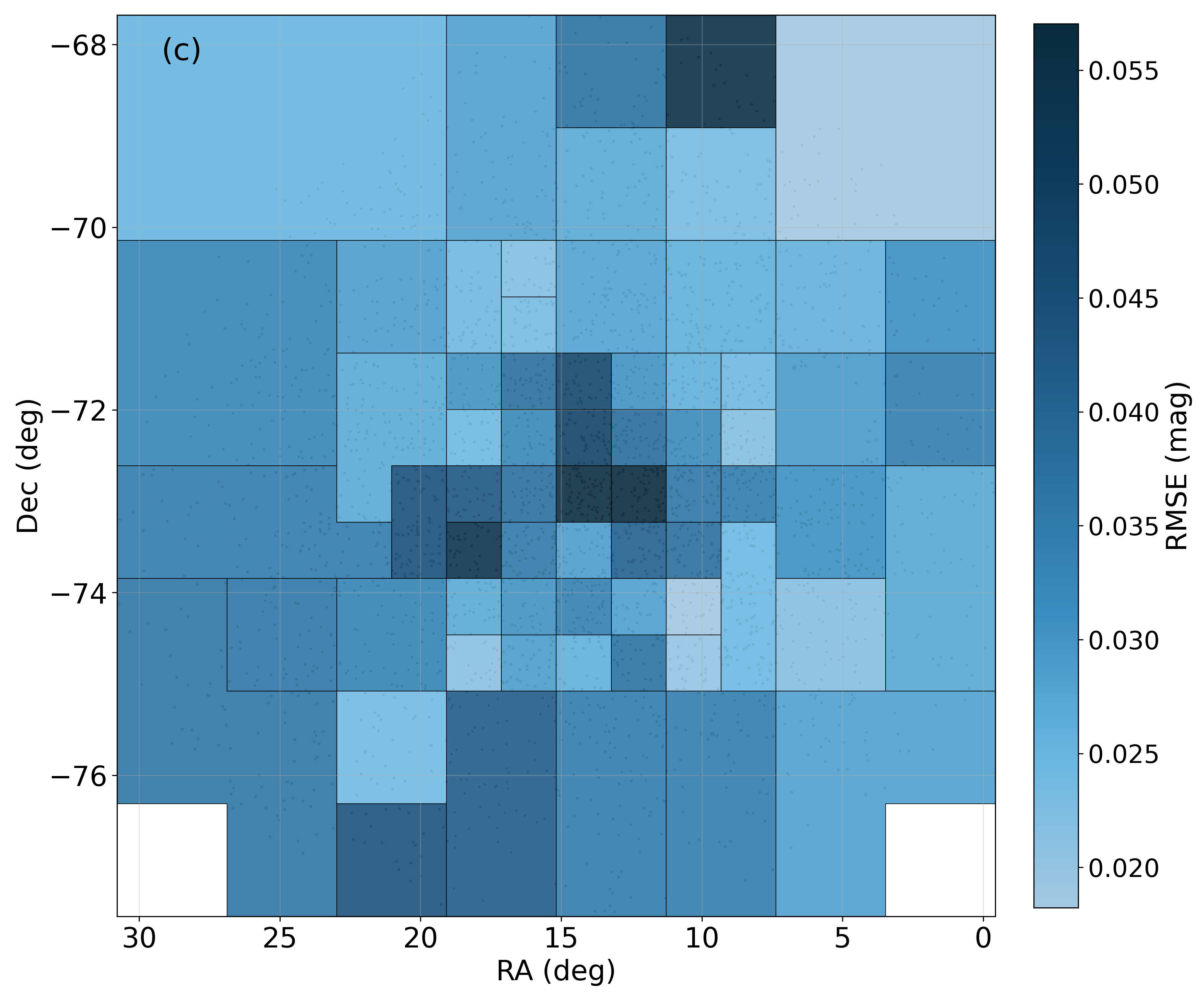}
  
  \vspace{0.2cm}
  
  \caption{Overview of the spatial subdivision and fitting performance of the SMC reddening map. Panel (a): Spatial partitions of the SMC, with each subregion labeled by its ID in red. Panel (b): Reddening gradients derived from the linear reddening–distance fits, color-coded by gradient magnitude. Star symbols mark representative star-forming regions, labeled according to the catalog of \citet{Henize1956}. Panel (c): RMSE of the fits in each partition, with darker colors indicating larger residuals.}
  \label{fig:SMC_fit_result}
\end{figure*}

\section{Results and Discussion}\label{sec:cali}

\subsection{The 3D reddening maps of MCs}

\begin{figure*}[!htbp]
  \centering
  \includegraphics[width=0.98\textwidth,keepaspectratio]{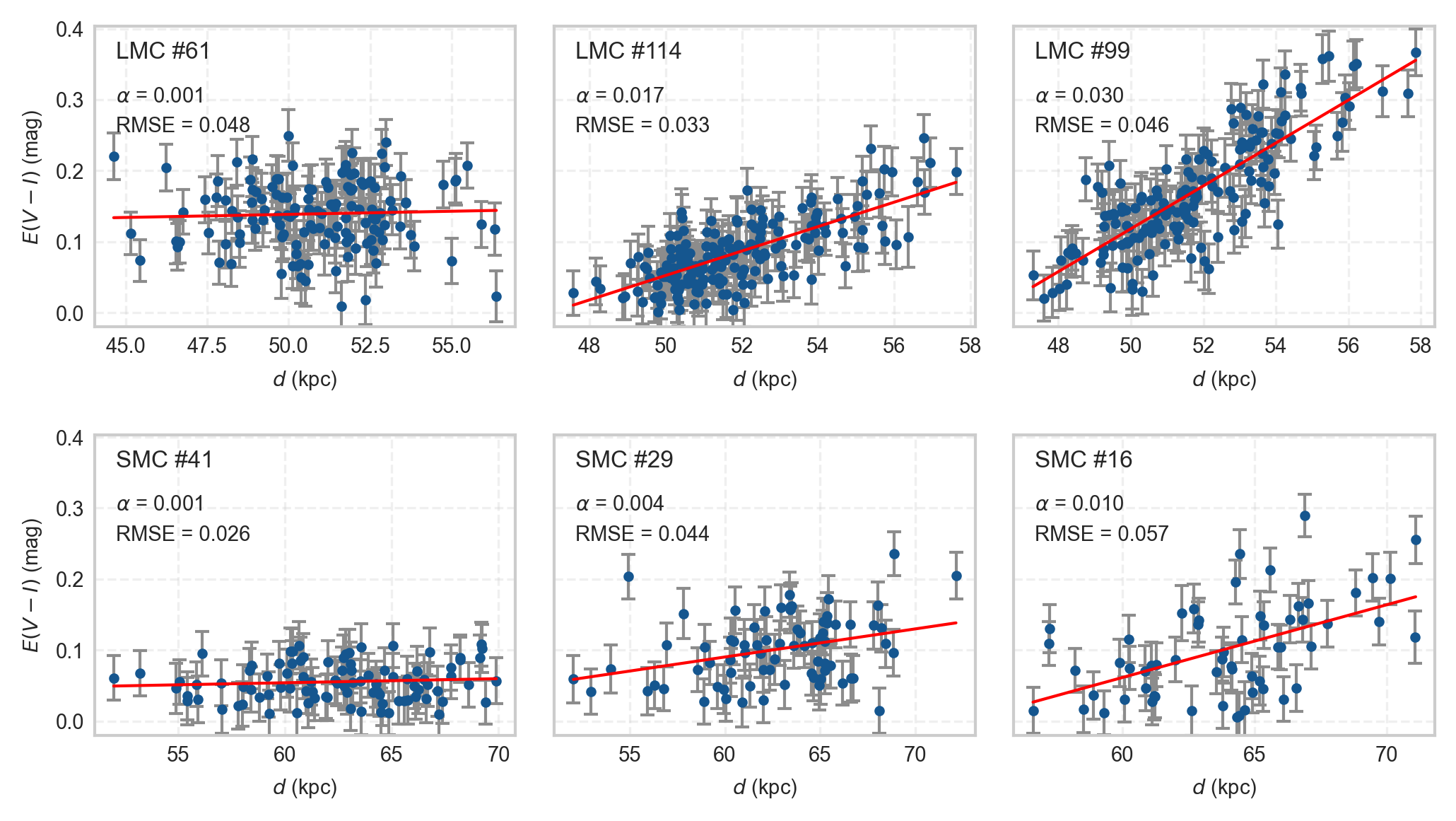}
  \caption{Examples of reddening--distance relations spanning different gradient strengths in selected partitions of the LMC (top row) and SMC (bottom row). Each panel shows the measured $E(V-I)$ values with error bars and the best-fit linear relation (red line) for a single partition, with the partition ID, slope ($\alpha$) and RMSE.}
  \label{fig:evi_d_relation}
\end{figure*}

Based on the \textbf{MC\_FINAL} sample and the adaptive spatial partitioning method, we constructed 205 partitions for the LMC and 67 partitions for the SMC. Each partition corresponds to a reddening–distance relation derived through iterative fitting until convergence. Figures~\ref{fig:LMC_fit_result} and \ref{fig:SMC_fit_result} present, respectively for the LMC and SMC, the partition ID map (panel (a)), the reddening gradient $\alpha$ map (panel (b)), and the RMSE map (panel (c)) of the final fitting results. Examples of reddening–distance relations spanning different gradient strengths are provided in Figure~\ref{fig:evi_d_relation}, and the relations for all partitions are provided in the Zenodo\footnote{\label{fn:zenodo}\href{https://doi.org/10.5281/zenodo.17717895}{10.5281/zenodo.17717895}}.

In the LMC partition map, smaller partitions are found in the central region (achieving a highest resolution of $\sim 37'\times15'$), consistent with the higher density of available stars there. Most of these central partitions exhibit reasonable positive reddening gradients and distinct reddening levels (as partition 99, 114 displayed in Figure~\ref{fig:evi_d_relation}), although a few show slightly larger fitting uncertainties, which may reflect dust clumping in dense environments, as discussed in Section~\ref{sec:caveats}. In contrast, the peripheral regions are divided into larger partitions due to the sparse spatial distribution of the sample. In these outer areas, where the line of sight intersects less dust, the increase in $E(V-I)$ tends toward zero or remains relatively small. Combined with the limited precision of $E(V-I)$ measurements and the larger partition size, a few outer partitions exhibit slightly negative reddening gradients (with the minimum not exceeding $-0.005\,\mathrm{mag\cdot kpc^{-1}}$), while most values lie slightly above but close to zero (as partition 61 displayed in Figure~\ref{fig:evi_d_relation}).The most pronounced reddening gradients occur in the vicinity of the 30 Doradus complex (centered at RA $=84.7^{\circ}$, Dec $=-69.1^{\circ}$). This extreme environment, illuminated by the R136 cluster, contains a porous interstellar medium with extended, high HI column densities and clumpy CO clouds structure \citep{Chevance2016,Clark2023}, collectively giving rise to the strong and spatially variable reddening observed in these partitions. Consistently, many of the representative star-forming regions marked in Figure~\ref{fig:LMC_fit_result} also exhibit comparatively elevated reddening gradients.

\begin{figure}[htbp]
  \centering
  \includegraphics[width=0.49\textwidth,keepaspectratio]{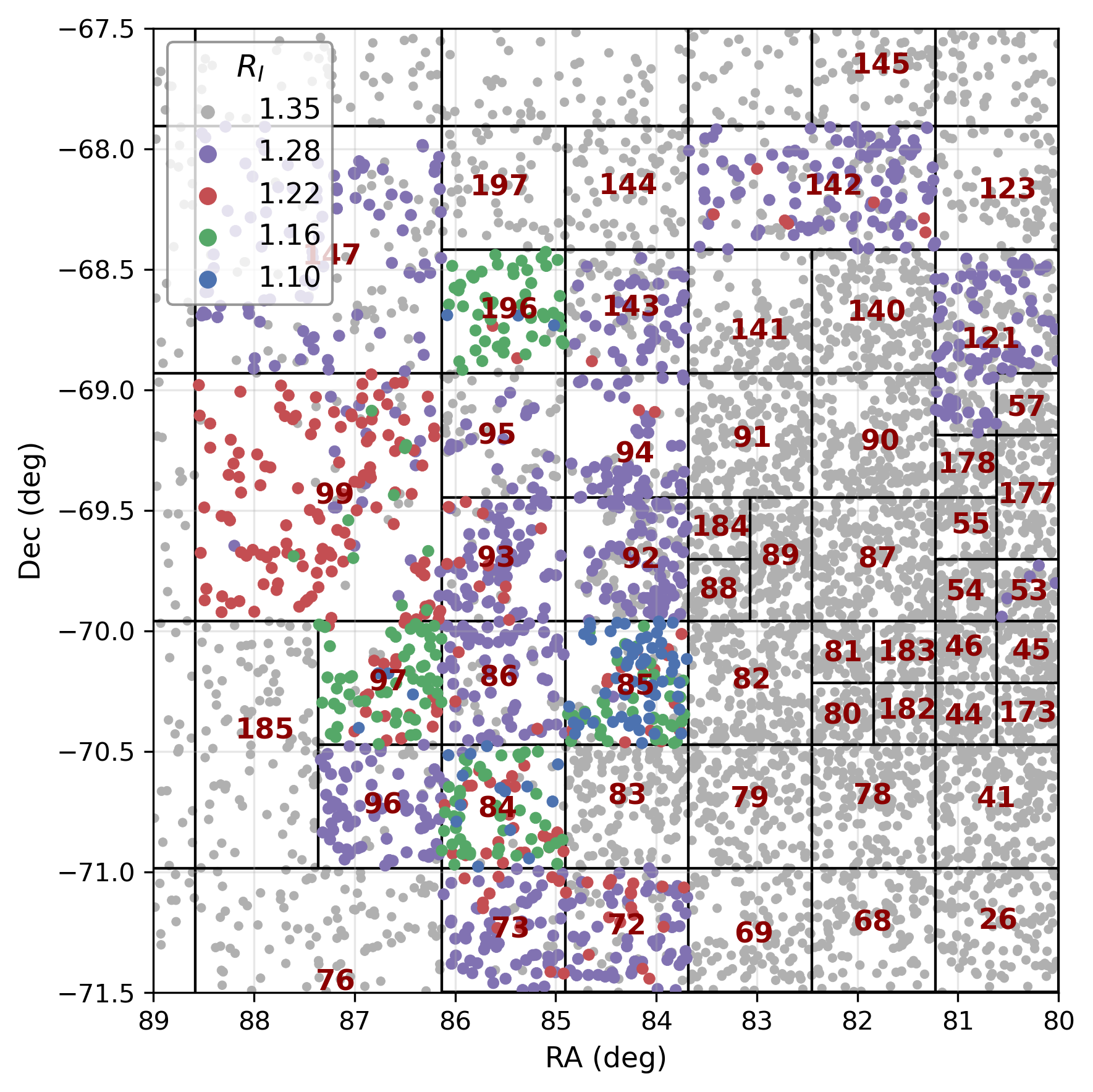}
  \caption{Zoomed-in view of the LMC region where the $R_I$ adjustment is applied. The plot shows the final $R_I$ values of individual sample stars, with different colors indicating different adjusted values: gray marks the initial $R_I$, while purple, red, green, and blue represent progressively smaller $R_I$ values.}
  \label{fig:R_I_zoomed_region}
\end{figure}

Another point worth noting concerns the $R_I$ adjustment results in the LMC. Figure~\ref{fig:R_I_zoomed_region} shows the distribution of $R_I$ values for the sample stars, where only regions with values lower than the initial assumption are displayed; all of these are located in the central area. Within each partition, the $R_I$ values are generally uniform or closely clustered, suggesting that the adjustments represent systematic partition-level variations rather than random star-to-star fluctuations. These lower $R_I$ values are primarily associated with dense, actively star-forming regions, indicating intrinsic modifications to the dust grain size distribution caused by grain growth and destruction processes, which in turn alter the local extinction law \citep{Zhang2025, Huang2025}. We further compare the relative $R_I$ variations with the $R_V$ map provided by \citet{Zhang2025}. The notable similarity between the two suggests that our adaptive $R_I$ adjustment approach may offer a new way to constrain extinction coefficients, thereby providing additional insight into the physical characteristics of the interstellar medium.

For the SMC, the partition map is limited by the smaller number of available sample stars. Even in the central regions, the partitions are larger than those in the LMC (highest resolution $\sim 1.9^{\circ}\times0.6^{\circ}$). As a result, a greater number of partitions exhibit negative reddening gradients in outer regions as shown in Figure~\ref{fig:SMC_fit_result}, similar to the behavior seen in the outer regions of the LMC. Nevertheless, different locations around the central region still show distinct reddening levels, as illustrated for partitions 16 and 29 in Figure~\ref{fig:evi_d_relation}. The steepest reddening gradients predominantly occur within H II regions characterized by high star‐formation activity (marked in Figure~\ref{fig:SMC_fit_result}) and are associated with clusters such as NGC 346, NGC 249 and NGC 456 \citep{Reyes2015}. The reddening gradients in the SMC are roughly half of those measured in the LMC. This difference is consistent with independent observations showing that the SMC has significantly less dust mass, a higher gas-to-dust ratio, and a more diffuse dust distribution than the LMC \citep{Gordon2014}. Observational effects may introduce mild smoothing in the derived gradients. The larger distance to the SMC increases distance uncertainties, the smaller number of tracers reduces spatial sampling, and its substantial line-of-sight depth can dilute sharp reddening variations. However, these factors mainly affect small-scale structures and are unlikely to explain the systematic difference in gradient amplitudes between the two galaxies. 
The observed contrast therefore most likely reflects intrinsic differences in their ISM properties, including the lower metallicity, reduced dust production efficiency, and overall lower dust content of the SMC \citep{Roman-Duval2014}.

To facilitate the practical use of our maps, we have developed a dedicated Python query tool that provides reddening estimates based on the full set of partition parameters encoded in a GeoJSON file. For any given sky position (RA, Dec) and distance within the range of the MCs, the tool identifies the corresponding partition through fast geometric lookup and returns the $E(V-I)$ value and uncertainty. Both command-line and file-based batch queries are supported. For partitions exhibiting negative reddening gradients, the tool applies our recommended refinement by replacing the linear model with the weighted mean $E(V-I)$, and this adjustment is noted in a remark field. Detailed usage instructions are provided in the README, and all relevant files have been deposited on Zenodo\footref{fn:zenodo}.

Overall, we present the first three-dimensional, distance-resolved reddening maps of the Magellanic Clouds, offering a new observational perspective on their internal dust morphology. The LMC shows steeper reddening gradients in its central, actively star-forming regions, whereas the SMC exhibits generally flatter gradients, highlighting fundamentally different large-scale dust environments. In addition, $R_I$ values are systematically lower in the innermost high-extinction regions than in low-extinction areas, indicating spatial variations in the extinction law that likely arise from differences in local dust properties or evolutionary processes. Together, these results provide a region-dependent, three-dimensional characterization of dust on scales of hundreds of parsecs to a few kiloparsecs, placing new constraints on the distribution and physical conditions of the interstellar medium within the Magellanic Clouds.

\subsection{Caveats}\label{sec:caveats}

The three-dimensional reddening gradients derived in this work are subject to several assumptions and observational limitations that should be considered when interpreting the results. These caveats mainly affect the achievable spatial resolution and the level of detail, but do not alter the large-scale trends discussed above.

\subsubsection{Assumptions on extinction coefficients and distance calibration}

The extinction coefficient $R_I$ adopted in this work is converted from $R_V$ value calibrated using red supergiants and classical Cepheids \citep{Wang2023}, which represent the average extinction law in the Magellanic Clouds. Since the extinction coefficients directly enter the distance determination, uncertainties in the adopted $R_V$ values and in their conversion to $R_I$ may propagate into the inferred reddening–distance relations. First, possible spatial variations in $R_V$ among different partitions could introduce systematic offsets in the inferred reddening gradients. However, our adaptive adjustment of $R_I$ in high-extinction partitions indicates that its sensitivity to such variations is at the level of $\sim$5\%, corresponding to distance changes of no more than $\sim$1\% based on error propagation, and thus to at most a $\sim$1\% variation in the derived reddening gradients even in high-extinction regions.

Second, the calculation of $R_I$ depends on the adopted effective wavelength, which itself is influenced by the stellar spectral energy distribution, the filter transmission curve, and the amount of extinction \citep[e.g.,][]{Groenewegen2024}. Variations in the effective wavelength would modify the value of $R_I$, thereby propagating into the inferred distances and the derived reddening–distance slopes. However, because the extinction curve is relatively flat in the near-infrared $I$-band regime, the sensitivity to such shifts is limited. The resulting impact is therefore small and is expected to remain below the statistical uncertainties of the measured reddening gradients.

In addition, the adopted PMZ relation \citep{Prudil2024} sets the absolute distance scale of the RR Lyrae sample and therefore directly affects the normalization of the reddening–distance relation. However, the iterative and self-consistent fitting procedure employed here substantially reduces the dependence on this absolute scale. As a result, the relative reddening gradients, both within and between the galaxies, are only weakly sensitive to plausible calibration uncertainties.

\subsubsection{Spatial and distance-dependent incompleteness}

\begin{figure*}[!htbp]
  \centering
  \includegraphics[width=0.49\textwidth,keepaspectratio]{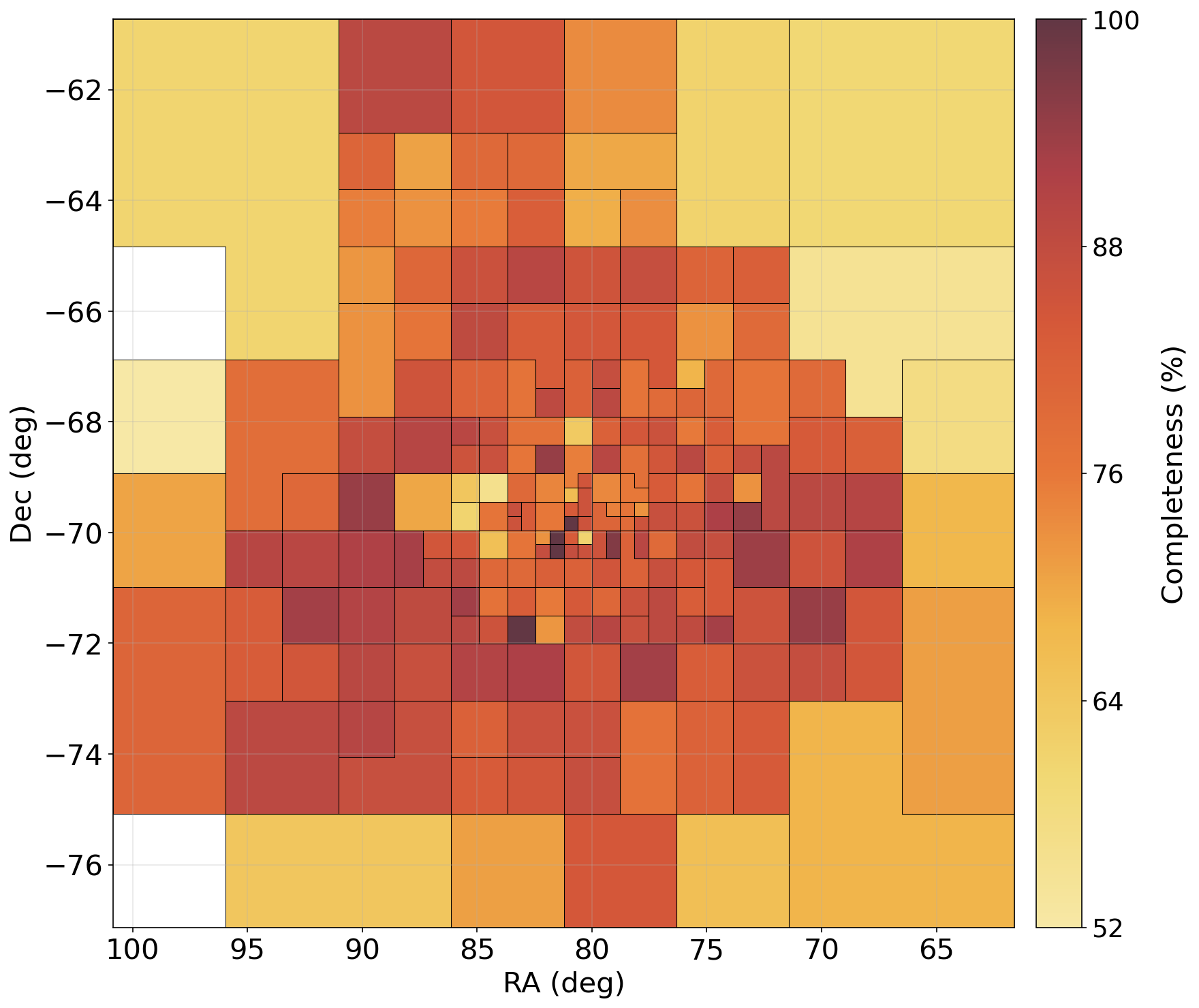}
  \hfill
 \includegraphics[width=0.49\textwidth,keepaspectratio]{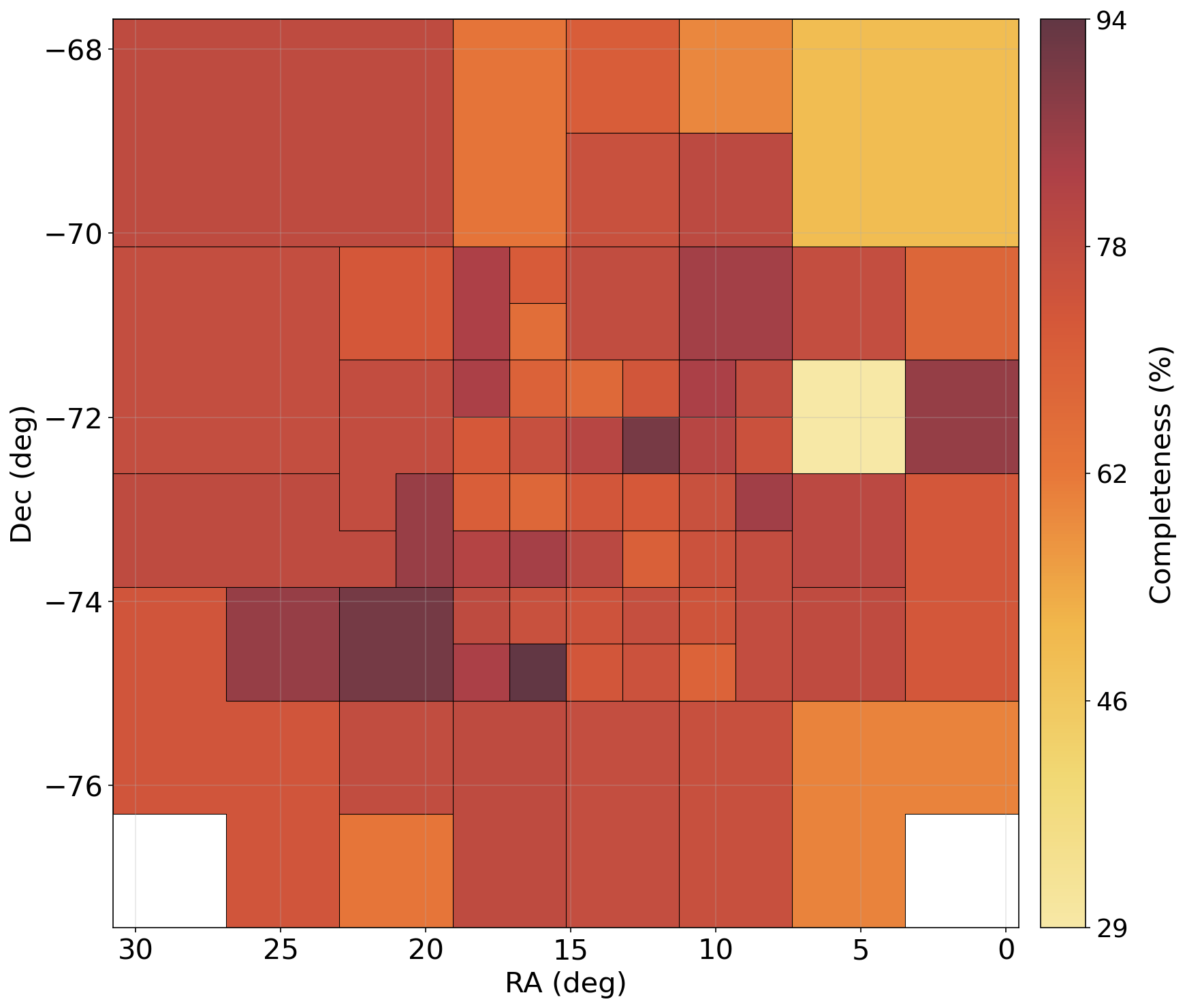}
  
  \vspace{0.2cm}
  \caption{Spatial completeness maps of the OGLE-based RR Lyrae sample. The left and right panels show the completeness in the LMC and SMC, respectively, defined relative to Gaia DR3 RR Lyrae members from the SOS pipeline. Colors indicate the completeness within each spatial partition.}
  \label{fig:Completeness of our OGLE sample}
\end{figure*}

The RR Lyrae sample exhibits both spatial and line-of-sight incompleteness, which systematically limits the effective resolution of the derived reddening maps. To quantify this effect, we compared our OGLE-based sample (after quality cuts) with Magellanic Cloud RR Lyrae members from the Gaia DR3 catalog processed through the Specific Object Study (SOS) pipeline \citep{Clementini2023}, adopting the completeness estimation framework from previous studies \citep{Mateu2020,Mateu2024}. As shown in Figure~\ref{fig:Completeness of our OGLE sample}, most partitions exhibit a typical completeness of approximately 80\% both in LMC and SMC. The lowest completeness in the SMC (partition 62) arises from the absence of $E(V-I)$ values required to compute the initial distances $d_\mathrm{ini}$, as this region is excluded in the reddening map of \citet{Skowron2021} due to contamination from the globular cluster 47~Tucanae. In contrast, a small number of highly crowded and high-extinction regions in the LMC, such as those near 30~Doradus (also evident in Figure~\ref{fig:R_I_zoomed_region}), show reduced completeness down to $\sim$60\%. This reduction is primarily driven by the initial distance and distance-uncertainty cuts applied in our analysis.

These quality cuts are necessary to remove low signal-to-noise measurements and obvious outliers, and to mitigate the impact of photometric blending in crowded fields, which can otherwise produce non-physical structures in reddening maps \citep{Jacyszyn2017,Cusano2021}. In regions with reduced tracer density, the adaptive partitioning scheme naturally produces larger spatial cells, such that the inferred reddening gradients represent spatially averaged values. As a consequence, smaller-scale dust structures and locally steeper gradients may be smoothed in the final maps.

In addition to these spatial limitations, line-of-sight incompleteness further blurs the reddening–distance relation. In an idealized scenario where dust is concentrated near the disk mid‑plane, one would expect a relatively steep reddening increase in front of the main dust layer, followed by a much flatter rise once the line of sight passes beyond the disk. Detecting such a break in slope could provide valuable constraints on the vertical dust distribution. In our data, however, we do not identify statistically significant transitions in the reddening gradients across any partition. Several observational effects conspire to smooth intrinsically sharp features along the line of sight. First, the survey detection limit makes the sample progressively more incomplete on the far side of each galaxy. Second, the uncertainty of single‑star distance estimates (median uncertainty of $\sim$7.2\%, corresponding to $\sim$4\,kpc at 50\,kpc) further smears any abrupt changes. Together with selection biases discussed above, these effects largely wash out the expected signature of a geometrically thin dust layer.

Consequently, the observed near‑linear increase in reddening should be interpreted as a large‑scale, averaged trend rather than a finely resolved vertical profile. In the absence of statistically significant evidence for discrete breaks or changes in slope, a linear reddening–distance model represents the most conservative and data‑supported description of the reddening variation. Introducing more complex functional forms would risk over‑interpreting features that are not robustly constrained by the current observations.

\subsubsection{Dust clumping and unresolved structure}

Dust in the MCs exhibits pronounced clumpy and filamentary structure on arcminute scales, as revealed by Herschel dust emission maps \citep{Kim2010} and high-resolution reddening map studies \citep{Gorski2020}. Such small-scale inhomogeneities are particularly prominent in dense and actively star-forming regions.

In our analysis, the adaptive spatial partitions typically span several tens of arcminutes and may therefore encompass multiple dust clumps and low-density voids. Structures on smaller angular scales cannot be individually resolved and are consequently averaged within each partition. The fitted reddening–distance slope in each region thus represents a large-scale estimate of the mean dust distribution, rather than the detailed internal structure of individual clouds.

Crucially, small-scale dust clumps introduce additional line-of-sight extinction variations within each partition. Because these fluctuations increase the local dispersion rather than introducing a systematic trend with distance, they primarily enlarge the residual scatter of the fit without significantly altering the inferred reddening–distance slope. In high-extinction regions, the RMSE shown in Figure~\ref{fig:LMC_fit_result} (panel c) exceeds the baseline level of roughly 0.04\,mag measured in low-extinction regions, where the dispersion is dominated by measurement and intrinsic uncertainties. The excess above this baseline may reflect an additional variance component that could plausibly arise from unresolved dust clumping, corresponding in quadrature to approximately 0.03–0.04\,mag. Given that this level is comparable to the effective noise floor of the analysis, it remains only marginally distinguishable and does not indicate a clearly resolved small-scale dust structure.

In summary, the maps robustly trace the large-scale dust distribution in the Magellanic Clouds, although the absolute amplitudes of the reddening gradients may be modestly reduced by the averaging effects described above.

\section{Summary}\label{sec:summ}

In this work, we constructed the first three-dimensional reddening maps of the Magellanic Clouds using OGLE-IV RRab stars as distance tracers. Intrinsic colors from a PAC relation and absolute magnitudes from a PMZ calibration provided color excesses and photometric distances for more than twenty thousand stars. After rigorous quality selections, these measurements enabled a reliable reconstruction of the distance-dependent reddening structure in both galaxies.

To capture spatial variations in reddening, we applied an adaptive quadtree partitioning of the sky and fitted linear reddening–distance relations within each cell. Distances were iteratively updated until self-consistency was reached. In high-extinction regions, an adaptive downward adjustment of $R_I$ was required to suppress oscillatory distance solutions, and the resulting spatial variations in $R_I$ likely reflect differences in local interstellar medium properties.

The final maps consist of 205 partitions in the LMC and 67 partitions in the SMC. The LMC shows steep reddening gradients in its central regions, whereas the SMC exhibits generally flatter profiles, consistent with their distinct global dust environments. While the inferred gradients are constrained in resolution by sample completeness and distance uncertainties, the maps robustly capture the large-scale, distance-dependent distribution of dust within both galaxies. All partition parameters and a query tool have been publicly released to support distance-dependent reddening corrections.

Looking forward, Gaia DR4 is expected to identify additional Magellanic RR Lyrae stars with high-cadence, high-precision photometry, enabling more accurate color excesses, metallicities, and distances. Such data will improve the resolution of future 3D MCs reddening maps and reveal finer substructure in the dust distribution. More precise reddening–distance relations may also uncover non-linear features that trace internal inhomogeneities in dense or patchy dust clouds.

\begin{acknowledgments}
Y.H. acknowledges the support from the National Science Foundation of China (NSFC grant No. 12422303), the Fundamental Research Funds for the Central Universities (grant Nos. 118900M122, E5EQ3301X2, and E4EQ3301X2), and the National Key R\&D Programme of China (grant No. 2023YFA1608303). 
This work is also supported by the National Key Basic R\&D Program of China via 2023YFA1608303 and the Strategic Priority Research Program of the Chinese Academy of Sciences (XDB0550103).
\end{acknowledgments}

\bibliographystyle{aasjournalv7}
\bibliography{ref}



\end{document}